\documentclass[aps,prb,twocolumn,reprint,noshowpacs,nofootinbib,noshowkeys,superscriptaddress,floatfix]{revtex4-2}
\usepackage[colorlinks=true,linktocpage=true,linkcolor=blue,citecolor=blue]{hyperref}
\usepackage{amsmath, amssymb}
\usepackage{multirow}
\usepackage{longtable}
\usepackage{comment}
\usepackage{color}
\usepackage{xspace}
\usepackage{cleveref}
\usepackage{graphicx, epsfig, ulem, dcolumn, bm, lipsum}
\usepackage[dvipsnames]{xcolor}
\usepackage{ragged2e}
\setcitestyle{numbers}
\setcitestyle{square,ulem}
\usepackage{siunitx}

\begin{document}


\title{Anomalous Electrical Transport in SnSe$_2$ Nanosheets: Role of Thickness and Surface Defect States}
\author{Aarti Lakhara}
 \affiliation{Department of Physics, Indian Institute of Technology Indore, Khandwa Road, Indore, Simrol, 453552, India}
\author{Lars Thole}
 \affiliation{Institut f\"{u}r Festk\"{o}rperphysik, Leibniz Universit\"{a}t Hannover, 30167 Hannover, Germany}
 \author{Rolf J. Haug}
 \email{haug@nano.uni-hannover.de}
 \affiliation{Institut f\"{u}r Festk\"{o}rperphysik, Leibniz Universit\"{a}t Hannover, 30167 Hannover, Germany}
 \author{P. A. Bhobe}
 \email{pbhobe@iiti.ac.in}
 \affiliation{Department of Physics, Indian Institute of Technology Indore, Khandwa Road, Indore, Simrol, 453552, India}

\date{\today}

\begin{abstract}

This work examines the influence of thickness on the electrical transport properties of mechanically exfoliated two-dimensional SnSe$_2$ nanosheets, derived from the bulk single crystal. Contrary to conventional trend observed in two-dimensional systems, we find a semiconducting to metallic resistivity behavior with decreasing thickness. The analysis of low-temperature conduction indicates an increased density of states at Fermi-level with decreasing thickness, which is further corroborated by gate bias dependent conductance measurement. The enhanced conductivity in thinner flake is attributed to the n-type doping arising from surface defect states. The presence and evolution of these defect states with thickness is probed by thickness-dependent room-temperature Raman spectroscopy. Our study provides insights into the thickness-dependent electronic transport mechanism of SnSe$_2$ and the crucial role of defect states in governing the observed conductivity behavior.   

\begin{description}
\item[keywords]

\end{description}
\end{abstract}

\maketitle

\section{\label{sec:level1}Introduction }
In MX$_2$-type two-dimensional (2D) layered materials, where M is a $d$- or $p$-block metal and X is a chalcogen (S, Se, Te), the structure consists of a metal layer covalently bonded to two adjacent chalcogen layers, forming a unit layer. These layers stack via weak van der Waals (vdW) interactions to form the bulk material. This layered architecture enables multiple degrees of freedom for tuning electronic properties of the MX$_2$ materials\cite{science}. For instance, modifying the stacking orientation has given rise to twistronics\cite{twistronics}, while vertical stacking of different 2D materials into heterostructures has led to novel phenomena such as secondary Dirac points\cite{dirac_emergence} and engineered band gaps\cite{hetro_dirac}. 

A key advantage of these layered materials has been the ease of exfoliation into multilayers and monolayers, attributed to the weak vdW forces between its layers. This structural tunability leads to significant tuning of its electronic properties, as monolayers often exhibit markedly different behavior than their multilayer counterparts\cite{Layer_controlled}. Notable among its various exotic properties is the quantum confinement effect that can induce a transition from indirect to direct band gaps with decreasing layer thickness\cite{quantum_confinement, PRL_MoS2}. Understanding how electronic properties evolve with varying number of layers is crucial for applications in twistronics and heterostructure-based devices. However, systematic experimental studies on layer-dependent electronic behavior remain limited and significantly lag behind advances on the computational front.

\begin{figure}
 \centering
    \includegraphics[width=1\linewidth]{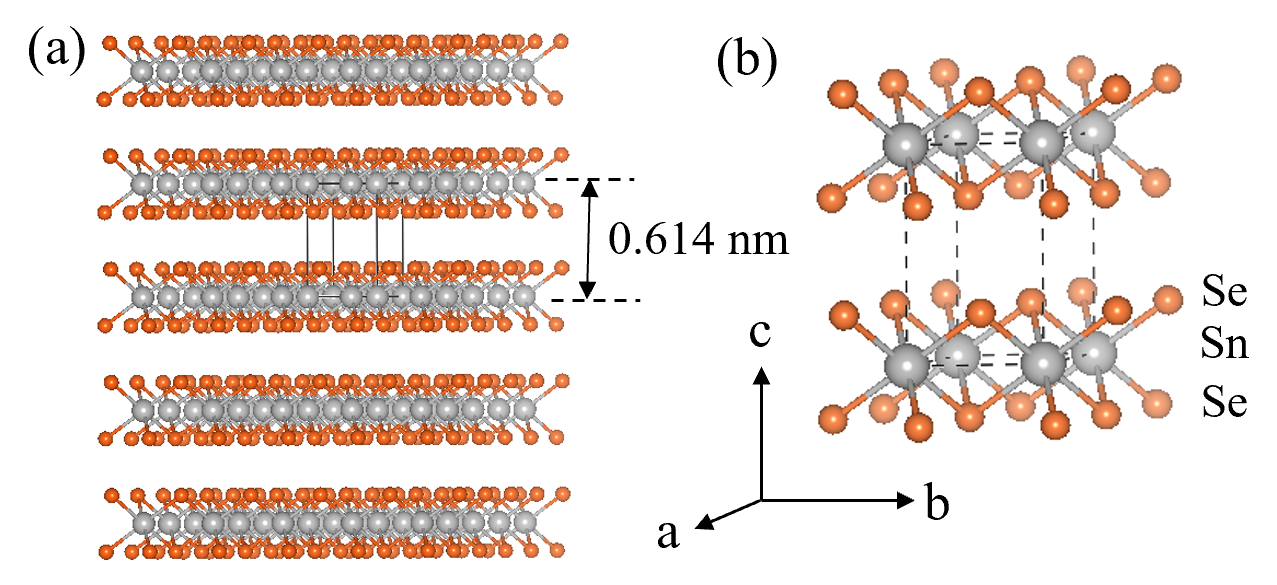}
\caption{\label{Strucutre_AFM} (a) Layered structure of SnSe$_2$ is shown with distance between two layer is 0.614 nm. (b) shows unit cell of SnSe$_2$.}   
\end{figure}

SnSe$_2$, a member of the metal dichalcogenide family, crystallizes in a hexagonal structure with Sn atoms sandwiched between Se layers. As shown in Fig.~\ref{Strucutre_AFM}(a), unit layers of SnSe$_2$ are stacked via vdW interactions to form the bulk crystal. SnSe$_2$ exhibits technologically relevant electronic properties and has been studied for various applications, including thermoelectric devices\cite{thermoelectric}, field-effect transistors \cite{su2013snse2, pei2016few}, and high-performance photodetectors \cite{zhou2015ultrathin}. Theoretically, SnSe$_2$ retains its indirect band gap nature even in the monolayer form, and the band gap increases with decreasing number of layers, from 1.07 eV in the bulk to 1.69 eV in the monolayer \cite{Layerdep2016prb}. The mechanism of electronic transport in SnSe$_2$ remain insufficiently explored. There are discrepancies in its reported temperature-dependent resistivity behavior. For example, Jianjun Ying \textit{et al}. observed semiconducting characteristics in bulk SnSe$_2$ single crystals \cite{PRL_SnSe2}. However, more recent studies on crystals grown via a modified Bridgman method have revealed that the in-plane resistivity increases with temperature above 110 K (metallic behavior), whereas in the low temperature range below 110 K, the resistivity decreases with rising temperature (semiconducting behavior). In contrast, out-of-plane resistivity increases continuously with decreasing temperature, indicating persistent semiconducting behavior\cite{Li_intercalated}. Transport properties also show strong thickness dependence. A 30 $\mu$m-thick crystal displayed semiconducting behavior below 100 K and metallic behavior above \cite{PhysRevMate_23}, while thinner flakes of 75 nm-thick crystal showed semiconducting behavior across the entire temperature range \cite{PhysRevMate_23}. Interestingly, an ultrathin flake with a thickness of 8.6 nm demonstrated metallic behavior between 78 K and 300 K \cite{guo2016field}. These findings indicate a strong thickness-dependent evolution of electronic transport in SnSe$_2$ though the underlying mechanisms remain unclear.

To address this gap, we systematically studied thickness-dependent electronic transport in SnSe$_2$. While bulk SnSe$_2$ single-crystal flake exhibits semiconducting behavior, we observe a transition to metallic transport when the flakes are thinned down to $\sim$ \qty{8}{nm}. This transition contradicts the theoretical prediction, which claims an increased band gap with a reduced thickness. Our analysis suggests that the anomalous metallic behavior in thinner samples is driven by donor-like surface defect states. The presence of these defect states is further supported by thickness-dependent Raman spectroscopy. This study provides a comprehensive understanding of the transport mechanism in SnSe$_2$ and highlights the role of surface defects in governing its electronic properties.

\begin{figure}
 \centering
    \includegraphics[width=1\linewidth]{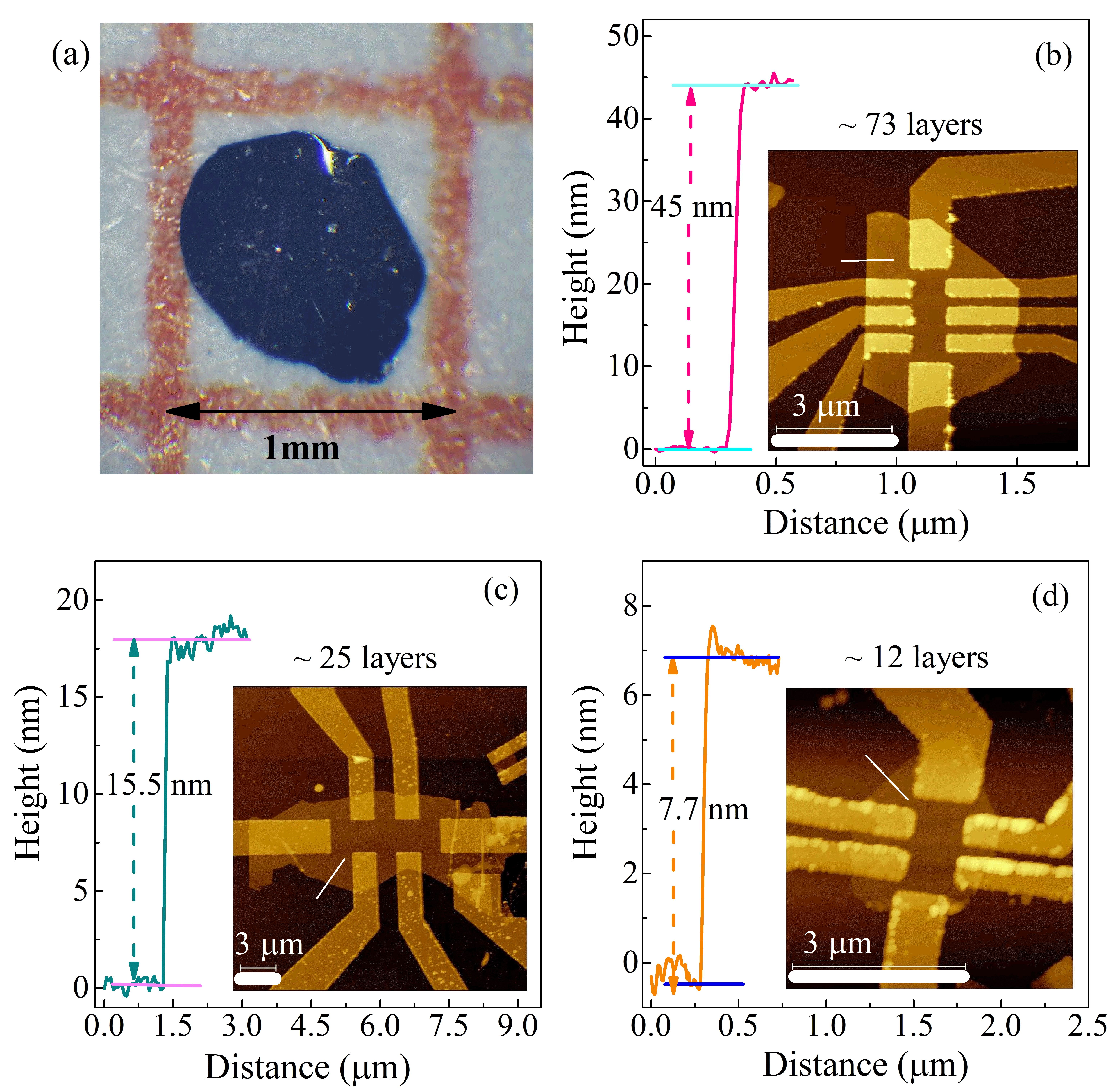}
\caption{\label{Cry_img_AFM} (a) is an optical image of bulk SnSe$_2$ single crystal with 1 mm scale bar. The height profile and AFM image of SnSe$_2$ flake thickness of (b) 45 nm, (c) 16 nm and (d) 8 nm (for simplicity the exact thickness values are round off and corresponding number of layers are also shown). The white line in each AFM image indicates where the AFM height profile is measured.}   
\end{figure}

\section{\label{sec:level2}Experimental details}
The bulk single crystals of SnSe$_2$ were synthesized using the chemical vapor transport (CVT) method. An optical image of a representative sample from the batch of synthesized crystals is shown in Fig.~\ref{Cry_img_AFM}(a). Thin flakes were obtained via standard mechanical exfoliation from the well-characterized bulk crystal and transferred onto a \qty{340}{nm} SiO$_2$ substrate. Electron beam lithography (EBL) was employed to fabricate alignment markers, electrodes, and bonding pads. Electrical contacts were made using Cr/Au with typical thickness of \qty{7}{nm}/ \qty{43}{nm}. To minimize surface oxidation, the exfoliated flakes were briefly exposed to air only during AFM measurement and during bonding of the device. At all other times, the samples were protected with a poly methyl methacrylate (PMMA) layer and stored in a nitrogen glove box or under vacuum.

For electrical characterization, three different SnSe$_2$ flakes on SiO$_2$/Si were selected by optical microscope. Their thicknesses were precisely measured using atomic force microscopy (AFM). The height profile of each thickness is shown in Fig.~\ref{Cry_img_AFM}(b)–(d), along with the AFM image with electrical contacts of three different thicknesses. The measured thicknesses were \qty{45}{nm}, \qty{15.5}{nm}, and \qty{7.7}{nm}, which are rounded to \qty{45}{nm}, \qty{16}{nm}, and \qty{8}{nm} for simplicity. The corresponding number of layers turn out to be $\sim$ 73, 25 and 12 respectively. 
The Raman spectroscopy (LabRAM HR Evolution, Horiba Scientific’s Raman spectrometer) was performed with two different laser excitation 2.33 eV (532 nm) and 1.96 eV (633 nm).

\section{\label{sec:level3}Results}
\begin{figure}
 \centering
    \includegraphics[width=1\linewidth]{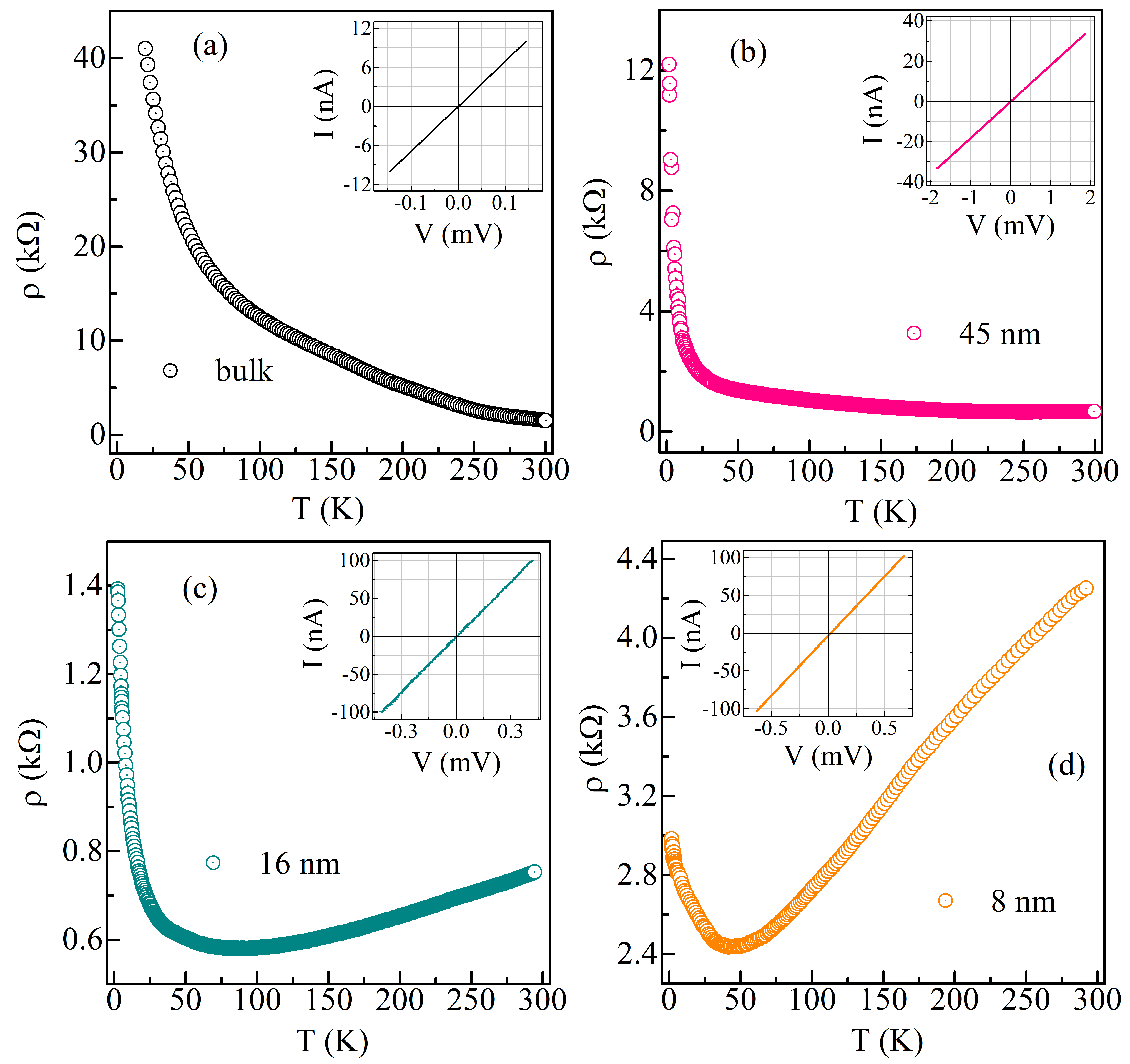}
\caption{\label{Rawdata} Temperature-dependent resistivity of (a) bulk single crystal, (b) 45 nm, (c) 16 nm, and (d) 8 nm flake; the insets shows linear I-V curve confirming the ohmic contact.}   
\end{figure}

To investigate the evolution of electronic properties with thickness in SnSe$_2$, we performed temperature-dependent sheet resistivity (hereafter referred to as resistivity) measurements on samples of varying thicknesses, including a bulk single crystal and exfoliated flakes with thicknesses of 45 nm, 16 nm, and 8 nm. The sheet resistivity was obtained using equation $\rho =\frac{W}{L}R$, where L and W are length and width of the channel and R is measured resistance.
Temperature-dependent resistivity measurements were carried out using a standard four-probe contact configuration. The insets of Fig.~\ref{Rawdata}(a)–(d) display linear I–V characteristics for each sample, confirming the formation of ohmic contacts.
As illustrated in the main panels of Fig.~\ref{Rawdata}(a)–(d), the resistivity behavior exhibits a pronounced dependence on sample thickness. The bulk SnSe$_2$ crystal demonstrates a typical semiconducting transport behavior across the entire temperature range. Similarly, the 45 nm thick flake shows semiconducting characteristics closely resembling those of the bulk crystal. In contrast, the 16 nm flake displays a crossover in transport behavior, it exhibits metallic resistivity (i.e., increasing resistivity with increasing temperature) above 100 K, while transitioning to semiconducting behavior below this temperature. Most notably, the 8 nm flake shows metallic behavior down to 41 K, below which the resistivity increases with decreasing temperature, indicating a reemergence of semiconducting behavior at low temperatures.

\begin{figure}
 \centering
    \includegraphics[width=0.9\linewidth]{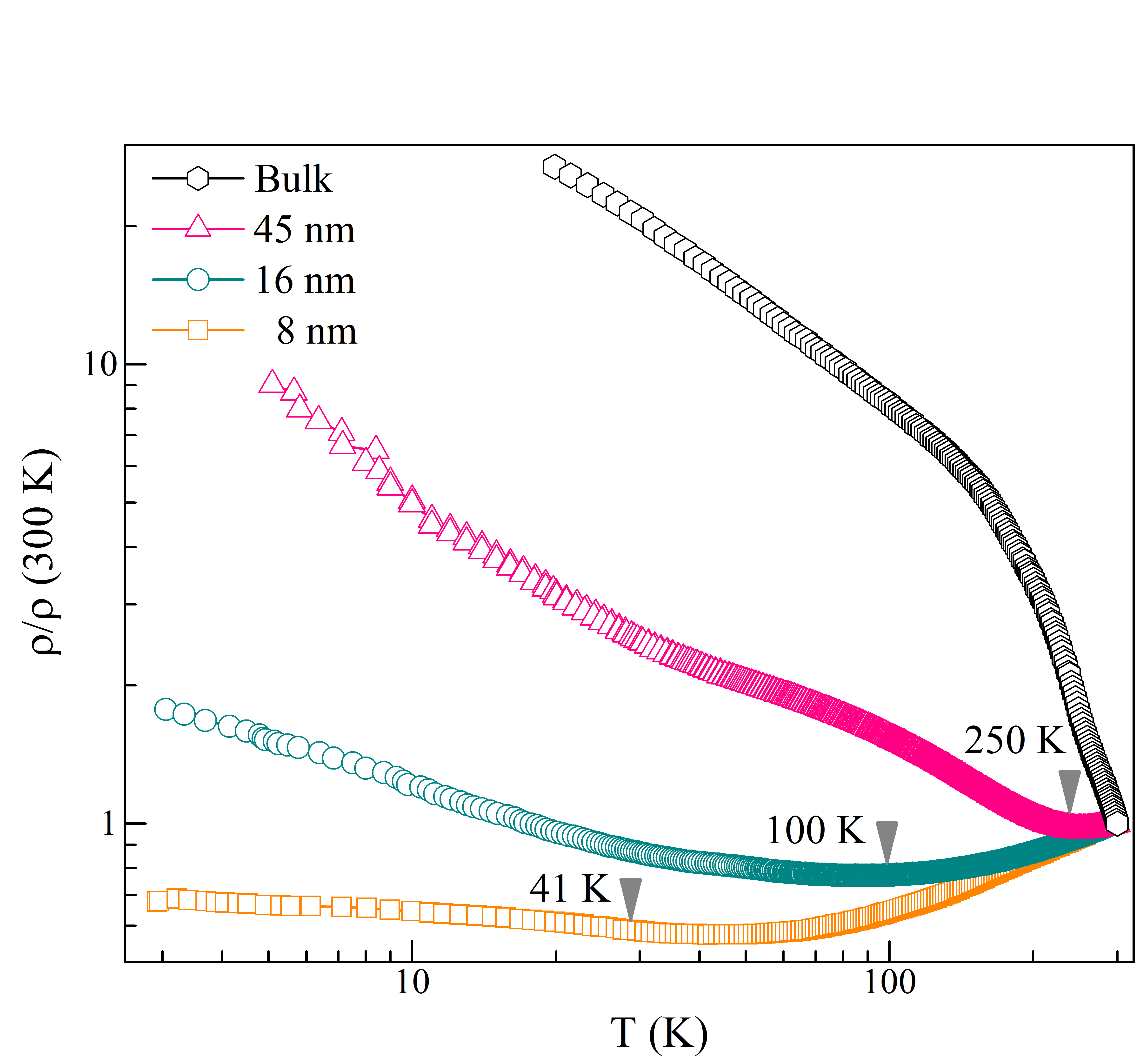}
\caption{\label{norm} Normalized temperature dependent resistivity of SnSe$_2$ bulk crystal and thin flakes.}   
\end{figure}

To more clearly distinguish the thickness-dependent resistivity behavior, the $\rho$ versus temperature curves were normalized to their respective resistivity values at 300 K, and both the x- and y-axes were plotted on a logarithmic scale, as shown in Fig.~\ref{norm}. This normalization highlights the systematic decrease in resistivity with decreasing thickness. The bulk SnSe$_2$ crystal maintains semiconducting behavior throughout the entire temperature range. The 45 nm flake also exhibits semiconducting characteristics similar to the bulk; however, a slight increase in resistivity is observed above 250 K with rising temperature, indicating a deviation from purely activated transport. The 16 nm flake displays metallic behavior above 100 K, transitioning to semiconducting behavior below this temperature. In the case of the 8 nm flake, metallic behavior persists down to 41 K, below which an upturn in resistivity is observed, a semiconducting behavior at low temperatures. The transition temperature between metallic and semiconducting transport regimes for each thickness was identified by the sign change in the temperature derivative of resistivity (d$\rho$/dT) and is marked with arrows in Fig.~\ref{norm}. Notably, this transition temperature decreases systematically with decreasing thickness, suggesting a modification in the electronic density of states as the material approaches the lower number of layers.

\begin{figure*}
\centering
 \makebox[0.7\textwidth]{\includegraphics[width=0.7\paperwidth]{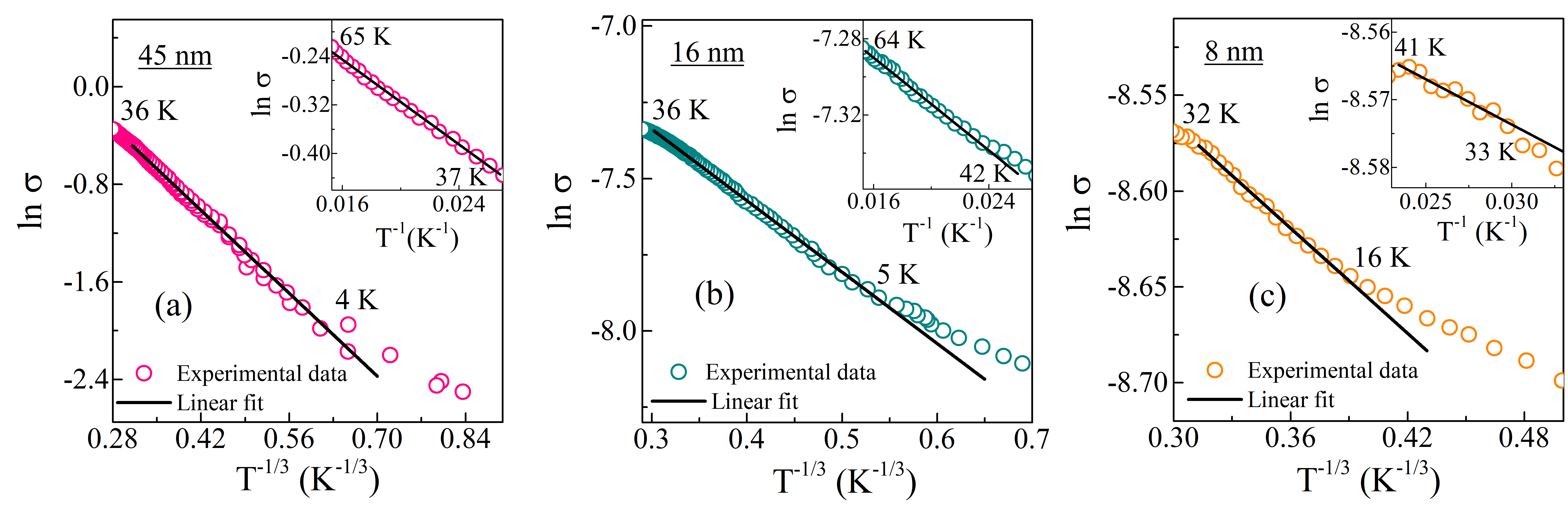}}

\caption{\label{Low_T_analysis}Low-temperature thickness-dependent resistivity analysis.  ln $\sigma$ as function of $T^{-1/3}$ for (a) bulk (b) 45 nm (c) 16 nm (d) 8 nm; Symbols represent experimental data, and the red line shows a linear fit, indicating a 2D VRH transport mechanism. The inset of (a)-(c) shows NNH mechanism for each thickness (see main text for details).} 

\end{figure*}

To gain deeper insight into the thickness-dependent transport mechanism, the low-temperature resistivity data were analyzed. An observed increase in resistivity at low temperatures suggests that carrier transport is dominated by hopping processes involving localized charge carriers. In the case of SnSe$_2$, this low-temperature transport behavior can be attributed to hopping conduction facilitated by gapped states induced by selenium (Se) vacancies as discussed in our previous study \cite{lakhara2025metal}.
Below 36 K, the transport mechanism is best described by two-dimensional variable range hopping (2D-VRH), wherein electrons hop between spatially distant localized states. The temperature dependence of conductivity in this regime follows the Mott VRH model, which is expressed by the equation:

\begin{equation}
    \sigma \propto exp[-(T_0/T)^{1/(d+1)}] 
\end{equation}  
where $T_0$ is the characteristic temperature, d is the dimension, and $\sigma$ is the conductivity. For 2-dimensional variable range hopping d = 2. Above this temperature, transport occur via the nearest neighbor hopping (NNH) between nearest localized states.  For the NNH model, the conductivity is given by $\sigma$ $\propto exp(-E_a/k_B T)$, where E$_a$ is the activation energy and $k_B$ is the Boltzmann constant.

Building on this understanding, we applied the aforementioned transport mechanisms to analyze the thickness-dependent behavior at low temperatures in nanosheets, as illustrated in Fig.~\ref{Low_T_analysis}. Figures~\ref{Low_T_analysis}(a), (b), and (c) clearly show that for the 45 nm, 16 nm and 8 nm samples, the logarithm of conductivity (ln $\sigma$) exhibits a linear dependence on $T^{-1/3}$  below 36 K. This linear trend is indicative of 2D-VRH conduction. However, for the 8 nm sample, the linearity weakens and is only observed in the narrower temperature range of 32 K to 16 K.
At even lower temperatures, where the 2D-VRH model begins to deviate, the transport mechanism may transition to Efros–Shklovskii (ES) VRH, which accounts for Coulomb interactions between localized states. Nevertheless, confirming this transition requires further experimental investigation at ultra-low temperatures.
The insets of Figs.~\ref{Low_T_analysis}(a)–(c) show that ln $\sigma$ varies linearly with $T^{-1}$ at higher temperatures, consistent with NNH conduction. The characteristic temperature $T_0$, extracted from the VRH fitting, is tabulated in Table~\ref{tab:table1} for each thickness. A decreasing trend in $T_0$ with reduced thickness suggests an increase in the density of states at the Fermi level as the system becomes thinner \cite{Dos_VRH}.
Additionally, the activation energies derived from NNH fitting can be found in Table~\ref{tab:table1}. Notably, 8 nm exhibit lower activation energies compared to 45 nm, indicating enhanced carrier transport in thinner samples.

\begin{table}
\caption{\label{tab:table1} The characteristic temperature, activation energy, electron-phonon scattering strength, residual resistivity and Debye temperature obtained from resistivity. The carrier concentration and mobility obtained from transfer curve at 1.6 K and corresponding to zero back-gate voltage.}
\begin{ruledtabular}
\begin{tabular}{cccc}
Fitting Parameter & 45 nm & 16 nm & 8 nm \\
\hline
T$_0$ (K)&115.02 & 12.85 & 0.76\\
E$_a$ (meV)&1.52 & 0.52 & 0.11\\
$B_{ele-ph}$ (k$\Omega$)&-&1.49&5.44 \\
$\rho(0)$ (k$\Omega$)&-&0.57 &2.36 \\
$\Theta_D$ (K)&-&271&126\\
n$_{2D}$ $\times 10^{12}$ (cm$^{-2}$) & 0.36 & 2.06 & 12.87\\

\end{tabular}
\end{ruledtabular}
\end{table}

\begin{figure}
 \centering
    \includegraphics[width=1\linewidth]{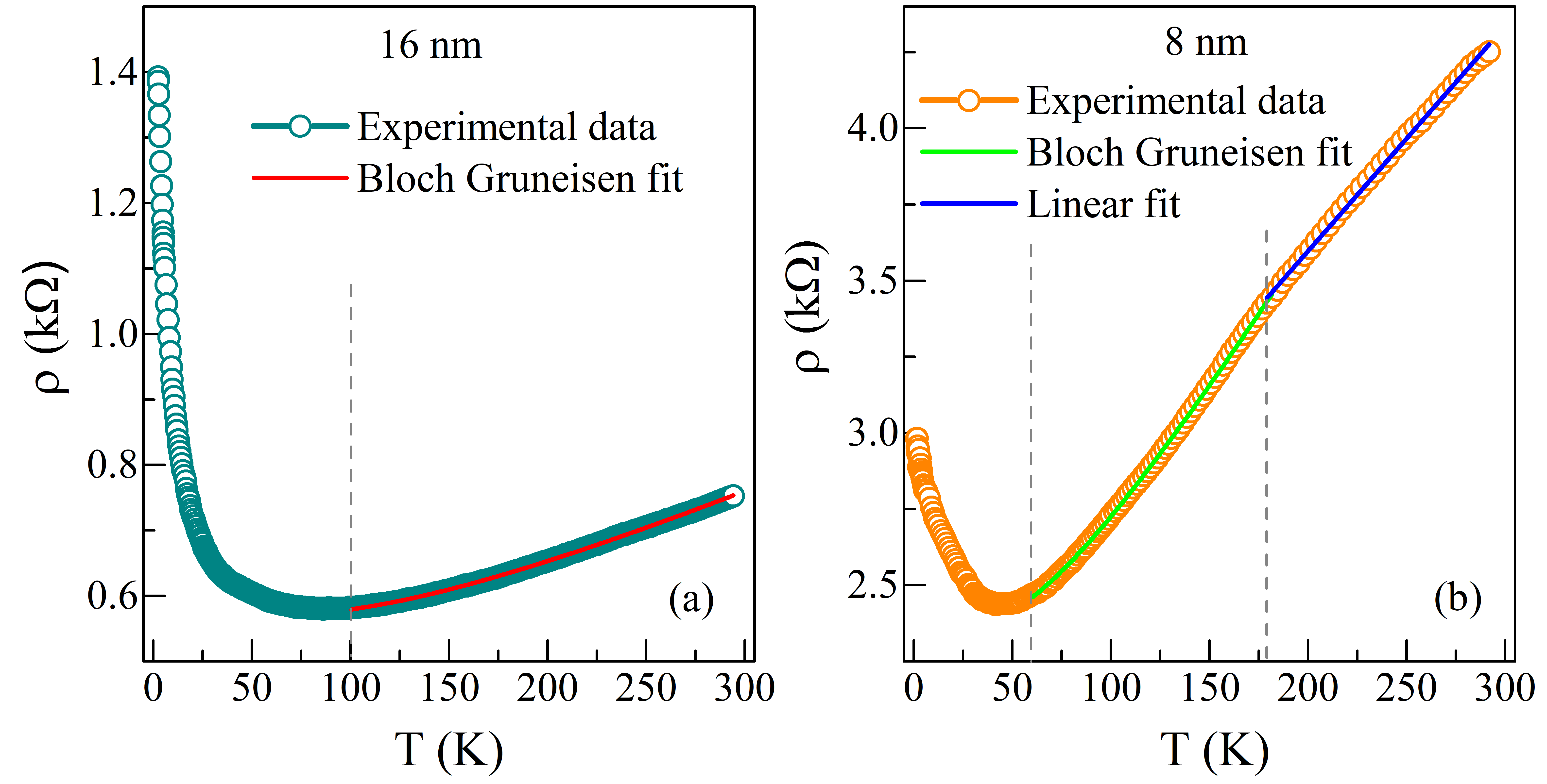}
\caption{\label{High_T_analysis} High-temperature experimental data of resistivity fitting in (a) 16 nm above 100 K and (b) in 8 nm above 60 K.}   
\end{figure}

This decreasing activation energy deviates from the conventional trend typically observed in 2D materials, where a reduction in thickness generally leads to an increase in activation energy due to quantum confinement effects \cite{Layerdep2016prb}. Theoretical studies on SnSe$_2$ also predict a similar trend, as the number of layers decreases, the activation energy should increase, respectively. Here, the thin SnSe$_2$ flakes under investigation were exfoliated from the same bulk crystal, ensuring consistency in both crystallographic quality and chemical composition. Additionally, SnSe$_2$ is known to retain its crystal structure even with decreasing thickness. Given this, the anomalous charge transport behavior observed in the thinner flakes is likely dominated by the surface states. One of the possible reasons for the creation of surface states could be the deselenization process that occurs during mechanical exfoliation. These states become increasingly influential as a result of the enhanced surface-to-volume ratio in ultrathin samples, which leads to the emergence of electronic states distinct from those in the bulk.

Next, to understand the high-temperature metal-like transport mechanism in 16 nm and 8 nm samples, we employed the Bloch-Grüneisen model, which describes the metallic conductivity in nonmagnetic systems due to electron-photon scattering. The Bloch-Grüneisen formula is given as follows:   
\begin{equation}
    \rho(T) = \rho(0) + \rho_{ele-ph}(T)
\end{equation}
\begin{equation}
\resizebox{0.9\hsize}{!}{$
\rho_{ele-ph}(T) = B_{ele-ph}\left(\frac{T}{\Theta_D}\right)^5\int_{0}^{\Theta_D/T} \frac{x^5}{(e^x-1)(1-e^{-x})}dx
$}
\end{equation}
Here $\rho(0)$ is residual resistivity and B$_{ele-ph}$ is electron-phonon scattering strength and $\Theta_D$ is Debye temperature.

As shown in Fig.~\ref{High_T_analysis} (a), the high-temperature data for the 16 nm samples above 100 K were well-fitted to the Bloch-Grüneisen model. For the 8 nm samples, the fit was accurate from 60 K to 180 K, as seen in Fig.~\ref{High_T_analysis} (b). Above 180 K, it was observed that the resistivity varies linearly with temperature. The fitting parameters of 16 nm and 8 nm can be found in Table~\ref{tab:table1}. The electron-phonon scattering strength and residual resistivity increases for 16 nm to 8 nm. This indicates the increased disorder and enhanced scattering in 8 nm flakes, which consequently lead to lowering of Debye temperature from 271 K to 126 K. The reported Debye temperature for bulk SnSe$_2$ is 190 K($\pm$ 5 K) \cite{thermoelectric,Li_intercalated}. An increase in the Debye temperature observed at a thickness of 16 nm relative to the bulk value suggests enhanced lattice stiffness resulting from thickness reduction. However, a further decrease in thickness leads to a reduction in the Debye temperature, indicating that the influence of disorder becomes more prominent than the effect of thickness-induced lattice stiffening. For completeness sake, we attempted to fit the high-temperature resistivity data for the 45 nm samples (above 250 K) using the Bloch-Grüneisen model; however, the fit was unsuccessful, likely due to its very limited temperature regime in the metallic phase.

\begin{figure}
 \centering
    \includegraphics[width=0.8\linewidth]{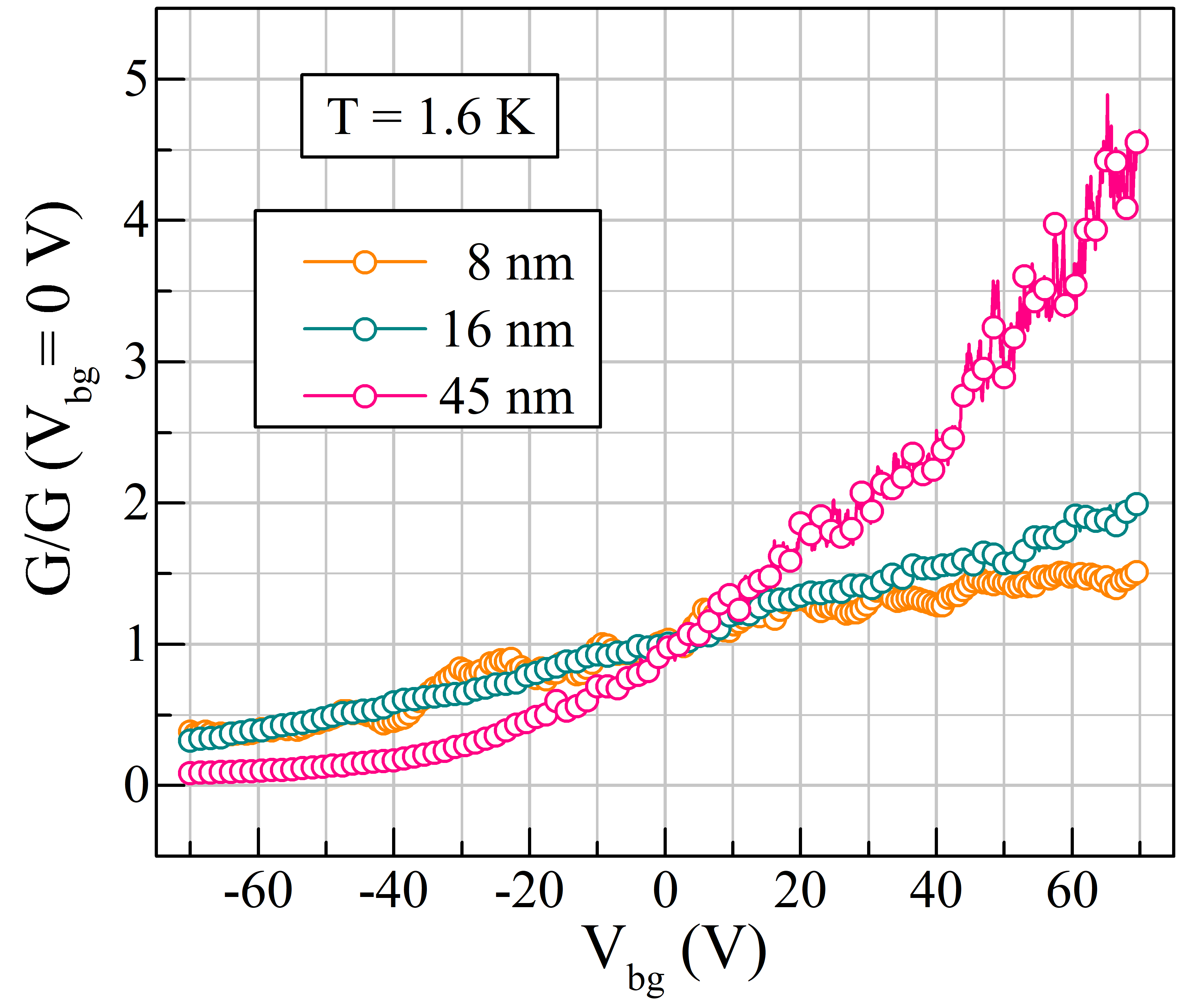}
\caption{\label{G_1.6K} Thickness-dependent transfer curve at 1.6 K.}   
\end{figure}

 Furthermore, in order to quantify enhanced surface states in thinner flake, we now turn our attention to the conductance measurement as a function of back-gate voltage for different thicknesses. In Fig.~\ref{G_1.6K}, we have plotted the normalized conductance with respect to zero back-gate voltage for each thickness at 1.6 K. This plot provides key differences in the properties of the different SnSe$_2$ thickness samples. Starting with the thinnest flake (8 nm), the conductance does not change significantly with the back-gate voltage, indicating efficient screening of the gate voltage by the high density of charge carriers in the sample, consistent with the metallic nature of this flake. Similar behavior is observed for a 16 nm thick flake with slightly high conductance in the positive back-gate voltage region, compared to 8 nm thick flake. In contrast the thickest sample (45 nm) exhibits a clear non-linear increase in conductance with an increase in back-gate voltage. This is the signature of a semiconductor. Bulk SnSe$_2$ possesses a finite band gap whereas the 8 nm thickness flake has a finite number of density of states around the fermi-level indicating significant contribution of defect states. As back-gate voltage is used to change carrier density in the channel flake, the conductance of a semiconductor is enhanced.  In contrast, the same back-gate voltage when applied to metals, which already have a high charge carrier density, the change induced by the back-gate voltage is weak.

 We have calculated the carrier density (n$_{2D}$) using the capacitance model in our devices, from the conductance curve at 1.6 K for each thickness and the obtained values can be found in Table~\ref{tab:table1}. It is evident that the carrier density increases with decreasing thickness, indicating n-type doping is higher in thinner flake. Using the DFT method, it is verified in literature  that introducing one Se vacancy in monolayer SnSe$_2$ will create two unbound electrons and consequently shift the fermi-level towards the conduction band \cite{lu2023unlocking}. By this argument, the Se vacancies (defect states) should be higher in thinner flake.

\begin{figure}
 \centering
    \includegraphics[width=1\linewidth]{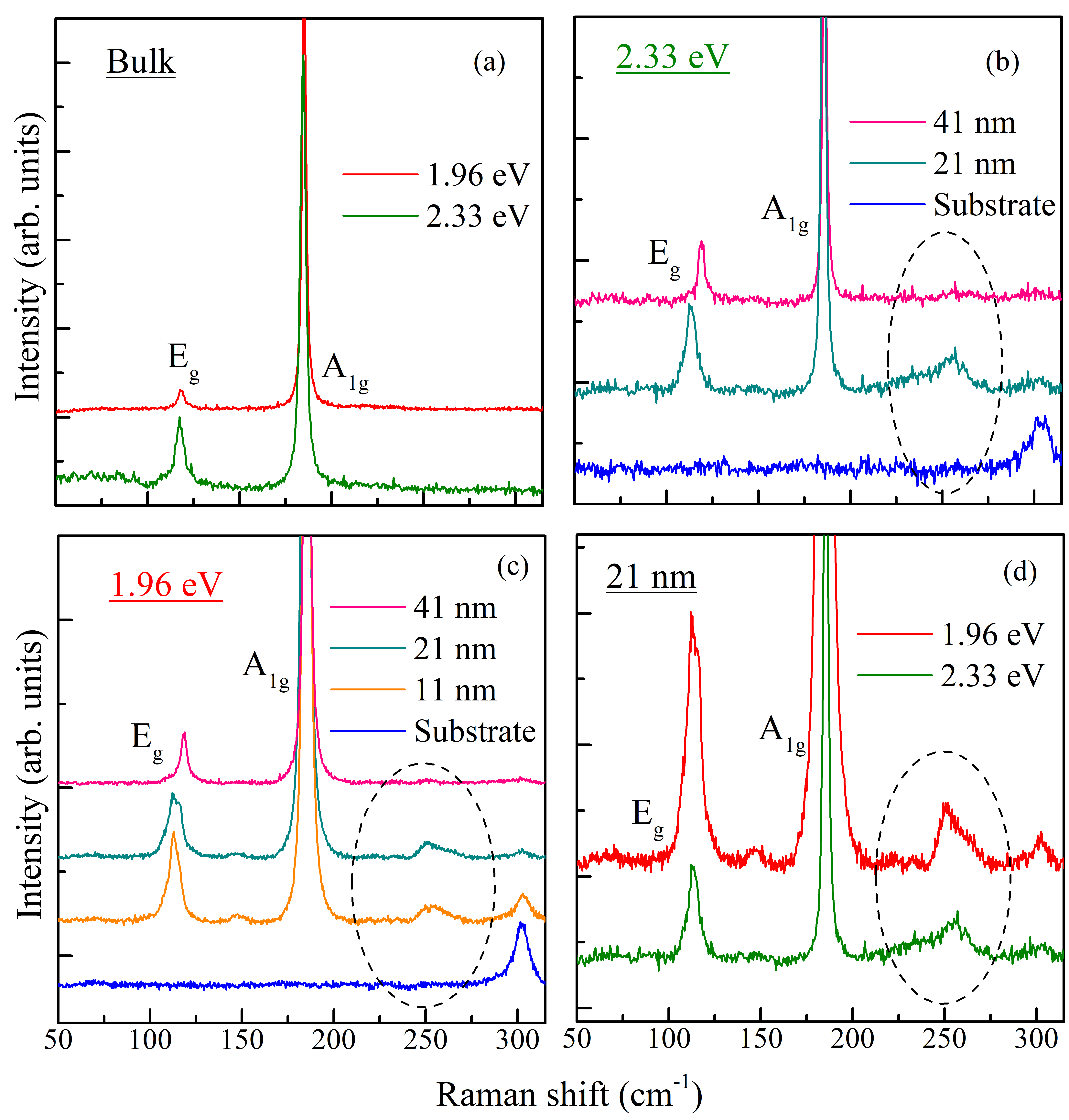}
\caption{\label{Raman} Thickness-dependent room-temperature Raman of SnSe$_2$ with 1.96 eV and 2.33 eV laser excitation. (a) Room-temperature Raman spectra of bulk single crystal. (b) Thickness-dependent Raman spectra at 2.33 eV laser excitation and spectra of substrate is also shown for comparison. (c) Thickness-dependent Raman spectra at 1.96 eV excitation. (d) Raman spectra of 21 nm thickness flake with two different laser excitation.}   
\end{figure}
To investigate the evolution of surface states and the conjecture of higher selenium vacancies in thinner SnSe$_2$ flakes, we performed room-temperature, thickness-dependent Raman spectroscopy, as shown in Fig.~\ref{Raman}. Spectra were recorded using two excitation energies, 1.96 eV and 2.33 eV. As shown in Fig.~\ref{Raman}(a), the Raman spectrum of the bulk SnSe$_2$ crystal exhibits two characteristic Raman-active modes, the in-plane vibration, E$g$ at $\sim$ 118 cm$^{-1}$, and the out-of-plane vibration, A${1g}$ at $\sim$ 185 cm$^{-1}$. The peak positions and the intensities of these modes are consistent with previous reports \cite{Raman_SnSe2}, and the absence of additional modes confirms the phase purity of the synthesized bulk crystals.

To study thickness effects, we examined flakes extracted from the same bulk crystal, having thicknesses of \qty{41}{nm}, \qty{21}{nm}, and \qty{11}{nm}. As shown in Fig.~\ref{Raman}(b) and (c), the \qty{41}{nm} flake displays only the E$g$ and A${1g}$ modes for both excitation energies, similar to the bulk crystal. Interestingly, upon reducing the thickness to \qty{21}{nm}, an additional high-frequency Raman mode appears around \qty{250} cm$^{-1}$ for both excitation energies. This high-frequency mode becomes more pronounced in the \qty{11}{nm} flake (measured with \qty{1.96}{eV} excitation), as shown in Fig.~\ref{Raman}(d). The high-frequency Raman modes are observed under both excitation energies (\qty{1.96}{eV} and \qty{2.33}{eV})(see Fig.~\ref{Raman} (d)), hence the possibility of any resonance mechanism giving rise to this feature, is ruled out. The emergence of this additional Raman mode in thinner flakes may likely be originating from selenium vacancies. A similar observation has previously been reported in other 2D materials like MoS$_2$ \cite{mignuzzi2015effect} and MoSe$_2$ \cite{maia2021defect}.  Unfortunately, Raman measurements for the \qty{11}{nm} flake using \qty{2.33}{eV} excitation could not be performed. However, it does not in any way influence the inferences drawn, as will be clear from further discussion. 

\begin{figure}
 \centering
    \includegraphics[width=1\linewidth]{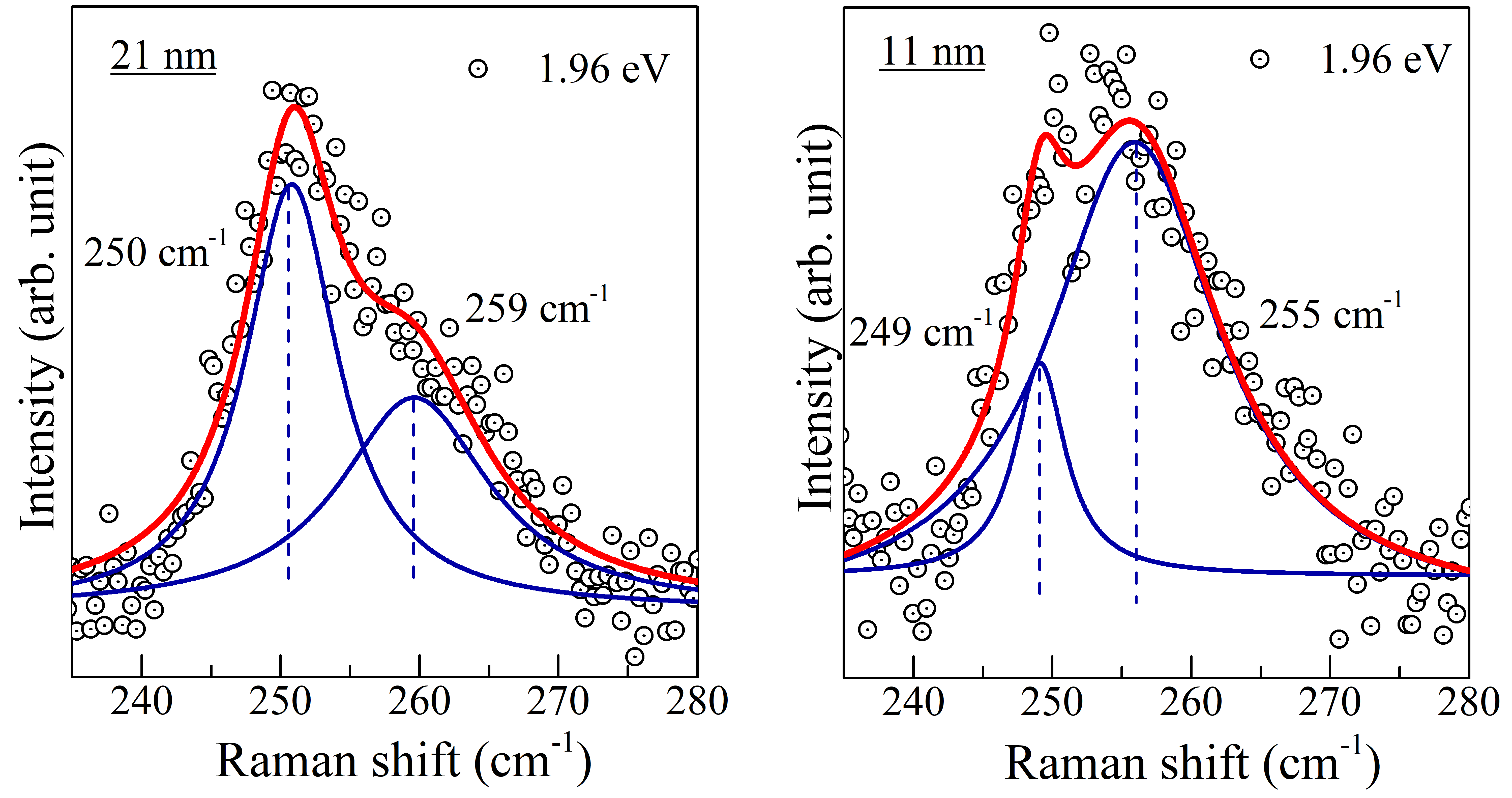}
\caption{\label{Raman_ana}  (a) and (b) shows Lorentzian fitting of defect induced peak in 21 nm and 11 nm flake.}   
\end{figure}

We examined the possibility of the additional features to be originating from IR-active phonons. Theoretically predicted IR-active transverse optical (TO) and longitudinal optical (LO) modes for bulk SnSe$_2$ occur at 241 cm$^{-1}$ and 248 cm$^{-1}$, respectively \cite{PhysRevB.14.1663_SnSe2_IR}. In Fig.~\ref{Raman_ana} (a) and (b) lorentzian fitting of high frequency Raman mode is shown for 21 nm and 11 nm respectively. Modes occurring at 250 cm$^{-1}$ for 21 nm and 249 cm$^{-1}$ for 11 nm may be associated with these IR-active phonons that are slightly red--shifted in the experimental scenario, and appear in thinner flake due to symmetry breaking. Additionally, a weak feature observed at 146 cm$^{-1}$ in thinner flakes can also be linked to an IR-active mode enabled by relaxed selection rules. However, a prominent contribution at high--frequency of 259 cm$^{-1}$ in 21 nm and 255 cm$^{-1}$ in 11 nm is observed, which cannot be solely explained by IR activity. As can be seen clearly from Fig.~\ref{Raman_ana} this high-frequency peak has at least two contributing modes, and one of them is due to the IR activity and the other is due to selenium vacancy. In addition, disorder-induced mode intensity increases with decreasing thickness implies that influence of disorder is higher in thinner flakes.

\section{\label{sec:level4} Discussion}
We investigated the thickness-dependent electronic transport properties of SnSe$_2$ and observed a transition from insulating to metallic behavior as the flake thickness decreases. This transition is attributed to an increased density of states near the Fermi level in thinner flakes, supported by back-gate voltage–dependent conductance measurements. This is in contrast to the theoretical calculation reported in Ref. \cite{Layerdep2016prb}, wherein SnSe$_2$ retains an indirect band gap even in the monolayer limit. Moreover, the band gap increases with decreasing number of layers due to reduced electrostatic screening and quantum confinement effects in few-layer systems.

The experimentally observed low-temperature transport and the corresponding analysis reveals that the activation energy associated with hopping conduction decreases with reduced thickness. This behavior is anomalous, as quantum confinement typically suppresses conductivity in thinner flakes. However, in our case, conductivity increases with decreasing thickness. One other case where such behavior is observed, is that of molybdenum chalcogenides, where surface electron accumulation dominates charge transport in thin films. This is in contrast to most semiconductors, where surface electron depletion is typically observed. To date, only a few bulk semiconductors such as, InAs \cite{noguchi1991intrinsic}, InN \cite{mahboob2004intrinsic, PhysRevLett.97.237601}, CdO \cite{piper2008observation, king2009unification}, and In$_2$O$_3$ \cite{king2008surface} have shown surface electron accumulation.

At the surface of a material, the perfect crystal periodicity typically breaks down, giving rise to surface states. In two-dimensional (2D) materials, the inherently high surface-to-volume ratio amplifies this effect. Moreover, mechanical exfoliation can further disturb the surface atomic layers, disrupting the periodic termination of unit cells. In the case of the SnSe$_2$ flakes under study, exfoliation results in imperfectly terminated surfaces, indicating a sudden break in lattice periodicity and the emergence of surface states.

\begin{figure}
 \centering
    \includegraphics[width=1\linewidth]{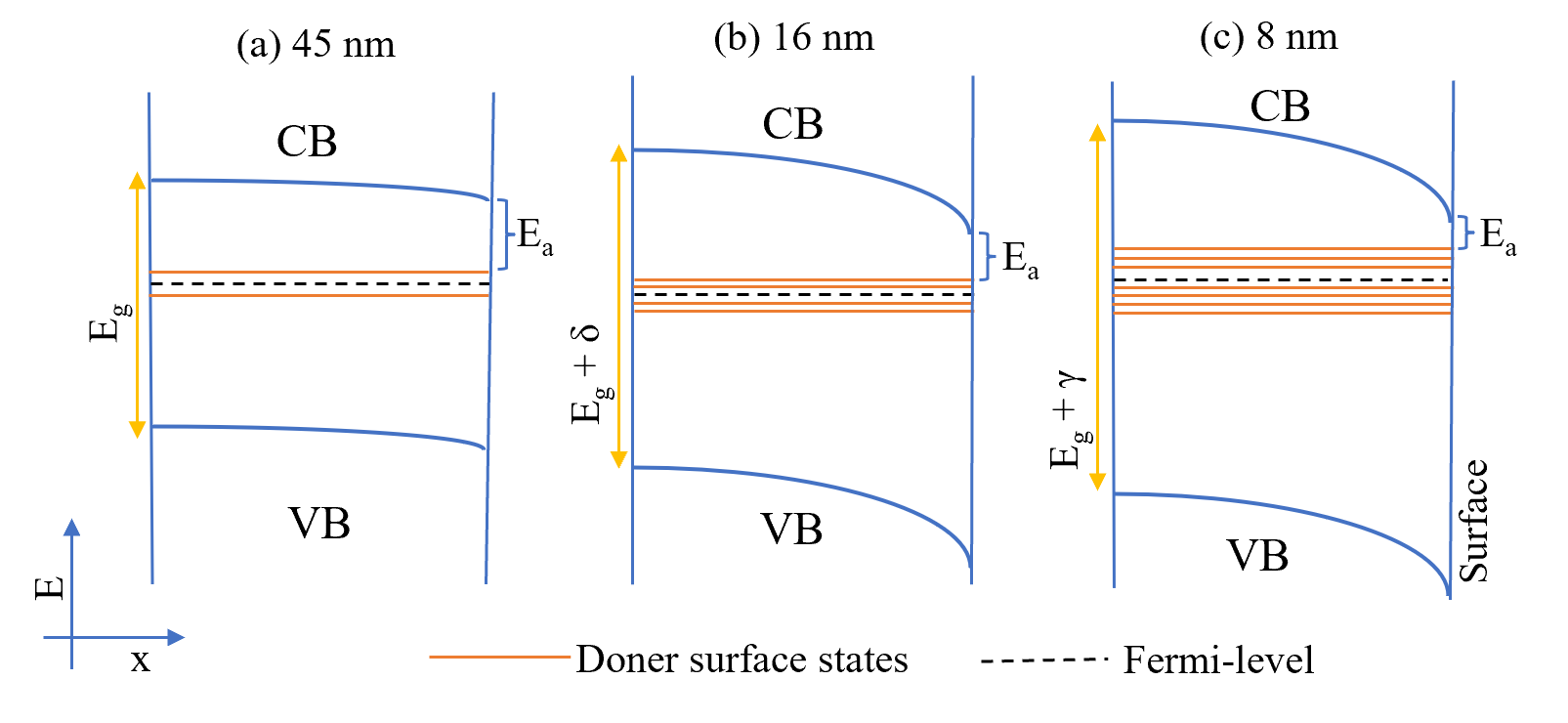}
\caption{\label{Schematic} Schematic diagram for the evolution of band structure with thickness (here CB refers to conduction band, VB refers to valence band and E$_a$ refers to activation energy). (a) is band structure picture for 45 nm flake where density of doner surface states are less near fermi-level and band gap is E$_g$. (b) is band structure picture for 16 nm where doner surface state density is higher compared to 45 nm and band gap is E$_g$ $+$ $\delta$ ($\delta>0$) along-with bending of bands towards surface. (c) is band picture of 8 nm where high doner surface density is observed near fermi-level and band gap is E$_g$ $+$ $\gamma$ ($\gamma>\delta>0$) along-with bending of bands towards surface.}   
\end{figure}

The origin of surface electron accumulation in SnSe$_2$ is linked particularly to selenium vacancies \cite{siao2018two, chang2021surface}. The DFT calculations \cite{lu2023unlocking} shows that the states caused by Se vacancies are situated near the conduction band edge, and introduce $n$-type doping in SnSe$_2$. As illustrated in Fig.~\ref{Schematic}, the effect of disorder primarily due to selenium (Se) vacancies, increases with decreasing thickness, driven by the enhanced surface-to-volume ratio. This leads to a higher density of states near the Fermi level in the 8 nm SnSe$_2$ flakes compared to those with thicknesses of 45 nm and 16 nm. Additionally, consistent with theoretical predictions the band gap (E$_g$) increment as the decrease in number of layers  is shown. The bending of the conduction band toward the surface becomes more pronounced in the 8 nm flakes, as a result of greater influence of donor-like surface states. This enhanced band bending and increased surface defect density collectively reduce the activation energy in the 8 nm samples relative to the thicker flakes.
Therefore, despite the widening of the band gap with decreasing thickness, the dominance of donor-like surface states in ultrathin SnSe$_2$ flakes induces a net metallic transport behavior. This underscores the critical role of surface defects in shaping the electronic properties of two-dimensional semiconductors.

\section{Conclusion}
We report on the thickness-dependent electronic transport properties of SnSe$_2$, revealing a transition from insulating to metallic transport behavior as the flake thickness decreases. This insulator-to-metal transition is attributed to an increased density of states near the Fermi level in thinner flakes, a phenomenon corroborated by back-gate voltage-dependent conductance measurements. The enhanced surface-to-volume ratio in ultra-thin SnSe$_2$, combined with intrinsic n-type doping arising from selenium vacancies, plays a crucial role in driving this metal-like behavior. The thickness-dependent Raman spectroscopy further supports this findings, showing the emergence of a selenium-vacancy–induced Raman mode exclusively in thin flakes. In particular, the observation of metallic transport in thinner SnSe$_2$ layers deviates from conventional expectations in two-dimensional materials, where changes in the number of layers typically affect the band gap without fundamentally altering the type of conduction. Our study emphasizes the significant influence of defect states on the transport properties of SnSe$_2$. 

\begin{acknowledgments}
This work was supported by German Academic Exchange Service (DAAD) funded project LUH-IIT Indore Partnership (2019-2023) under the ``A New Passage to India program". Aarti Lakhara acknowledges DST-INSPIRE (File: DST/INSPIRE/03/2019/001146/IF190704), New Delhi,
for providing the research fellowship. Work at Leibniz University Hannover was funded by the Deutsche Forschungsgemeinschaft (DFG, German Research Foundation) under Germany's Excellence Strategy - EXC 2123 Quantum Frontiers - 390837967 and within the Priority Program SPP 2244 '2DMP'. We also  acknowledge the Department of Physics, IIT Indore for Raman facility. We wish to acknowledge discussions on bandstructure with Tim Wehling and Lara Bremer from University of Hamburg, Germany.  
\end{acknowledgments}

\bibliography{manuscript}

\end{document}